\documentstyle[12pt]{article}
\newlength{\extraspace}
\newlength{\extraspaces}
\begin{document}

\addtolength{\baselineskip}{.7mm}
\thispagestyle{empty}
\begin{flushright}
{\tt hep-th/9801153} \\
December, 1997
\end{flushright}
\vspace{15mm}
\begin{center}
{\Large{\bf SUPERSYMMETRIC STRUCTURE OF SPACETIME AND MATTER \\[3mm]
-SUPERON QUINTET HYPOTHESIS-}} \\[30mm]
{\sc Kazunari SHIMA} \\[7mm]
{\it Laboratory of Physics, Saitama Institute of Technology} \\
{\it Okabe-machi, Saitama 369-02, Japan} \\[30mm]
{\bf Abstract}\\[5mm]
{\parbox{13cm}{\hspace{5mm}
A unified description of spacetime and matter at the Planck scale
is proposed by using the irreducible representation of N=10 extended
Super-Poincar\'e algebra,  where all matters and
all forces except the graviton are the supersymmetric composites
made of the fundamental objects with spin 1/2,
${\em superon}$ ${\em quintet}$.
All the local gauge interactions in GUTs  are investigated
systematically by using the superon diagrams.
The proton is stable and the flavor changing neutral current process is
suppressed in the superon pictures of GUTs.
The fundamental action of the superon model is proposed.
The characteristic predictions which  can be tested in the (coming)
high energy experiments  are discussed briefly.
}}
\end{center}
\vfill
\newpage

It seems generally accepted that the (local) supersymmetry (SUSY)[1] is
the most promissing notion for unifying all elementary particles
including the graviton within the framework of the local field theory.
However, as shown by Gell-Mann[2], SO(8) maximally extended supergravity
theory (SUGRA) is too small to accommodate all observed particles
as elementary fields.

On the contrary, at the risk of the local field theory at the moment,
it is interesting from the viewpoints of simplicity and beauty of Nature
to attempt the accomodation of all observed elementary particles in a single
irreducible representation of a certain group(algebra).
In the previous paper[3], by identifying the graviton as the Clifford
vacuum state( not necessarily the lowest energy state) of SO(N) extended
super-Poincar\'e algebra(SPA) we have studied
the irreducible representations of SO(N) SPA for the massless case.
And we have shown[3] that from only the group theoretical arguments
SO(N) SPA with N=10,11 and 12 may be relevant to the unified description
of matters and forces  and  N=10 stands out among them.
Because the assignment of quantum numbers adopted in Ref.[3]
to 10 supercharges $Q^{N}$ $(N=1,2,..10)$ of SO(10) SPA
is unique in order to realize
all observed quarks, leptons and gauge bosons as the low energy massless
states of the representations.
                                                                      \par
In this letter, after the brief review of Ref.[3] for the self-contained
arguments, we try to interpret the results of Ref.[3] from the viewpoints
of the internal structure of the quarks, leptons and gauge bosons except the
graviton.
We assume that at (above) the Planck energy scale Nature(spacetime and
matter) possesses SO(10) super-Poincar\'e symmetric structure and
that all of the massless irreducible representations of
SO(10) SPA reveal the structure.

In Ref.[3], by noting that 10 generators $Q^{N}$ of SO(10) SPA are the
fundamental represemtations of SO(10) internal symmetry  and that
$SO(10) \supset SU(5) \supset SU(3) \times SU(2) \times U(1)$
we have decomposed 10 generators $Q^{N}$ of SO(10) SPA as follows with
respect to SO(10) internal symmetry

\begin{eqnarray}
\underline{10} & = & \underline 5(Q^{N};N=1,2,..5)
+ \underline 5^{*} (Q^{N};N=6,7,..10) \nonumber \\
& = & \{(\underline 3, \underline 1;-{1 \over 3},-{1 \over 3},-{1 \over 3})
+
 (\underline 1, \underline 2;1, 0)\} +
 \{(\underline 3^{*}, \underline 1;{1 \over 3},{1 \over 3},{1 \over 3}) +
 (\underline 1, \underline 2^{*};-1,0)\},
\end{eqnarray}
where we have written ( SU(3), SU(2); electric charges ).
For massless case( $P_{\mu}P^{\mu}=0$ ) in order to see easily the helicity
contents
of the irreducible representation  we go to the little algebra, where
we can always choose the light-like frame $P_{\mu}=\epsilon(1,0,0,1)$.
In terms of two-component Weyl spinors, the little algebra for the
supercharges in this frame now becomes after suitable rescaling

\begin{equation}
\{ Q_{\alpha}^{M}, Q_{\beta}^{N} \}
=\{ \overline{Q}_{\dot\alpha}^{M}, \overline{Q}_{\dot\beta}^{N} \}=0, \quad
\{Q_{\alpha}^{M},\overline{Q}_{\dot\beta}^{N}\}
={\delta}_{{\alpha}1}{\delta}_{\dot\beta \dot 1}{\delta}^{MN},
\end{equation}
where $\alpha,\beta=1,2$ and $M,N=1,2,...10$.
As a cosequence of (2) the spinor charges
$Q_{1}^{M}$, $\overline{Q}_{\dot 1}^{M}$
satisfy the algebra of annihilation and creation operators respectively
and can be used to construct a 4-dimensional Fock space with positive
metric.
For the massless case, the Clifford vacuum $\mid\Omega(\lambda)\rangle$ is a
representation of the little group $E_{2}$ of a light-like vector,
i.e. a massless state of a given helicity $\pm\lambda$, if space inversion
is
considered.
We identify the graviton with the Clifford vacuum
$\mid\Omega(\lambda)\rangle$
(not necessarily the lowest energy state),
which satisfies

\begin{equation}
Q_{\alpha}^{M} \mid\Omega(\lambda)\rangle=0
\end{equation}

and build up a new state with helicity $(2-{n \over 2})$ by

\begin{equation}
\overline{Q}_{\dot{1}}^{M_{1}} \overline{Q}_{\dot{1}}^{M_{2}}....
\overline{Q}_{\dot{1}}^{M_{n}}\mid\Omega(\lambda)\rangle.
\end{equation}

(Note that the helicities of such states are  dtermined by the SO(10) SPA.)
These states given by the Clifford vacuum $\mid\Omega\rangle$ and
all states of (4) obtained from $\mid\Omega\rangle$ by multiplying with
every
possible product of the creation operators $Q_{\dot{1}}^{M}$
span an irreducible $2\cdot2^{10}$ dimensional representation of the
little algebra (2) of SO(10) SPA. It contains helicities up to $\pm{3}$,
if parity is included.
For a reference we show in the following explicitly all states of SO(10) SPA
and specify them by SO(10) dimension $\underline{d}$ and the helicity
$\lambda$, as $\underline{d}(\lambda)$:

\begin{eqnarray}
&& \Bigl[\underline{1}(+2), \underline{10}(+{3 \over 2}),
\underline{45}(+1), \underline{120}(+{1 \over 2}),
\underline{210}(0),
\underline{252}(-{1 \over 2}), \nonumber \\
&& \underline{210}(-1),
\underline{120}(-{3 \over 2}), \underline{45}(-2),
\underline{10}(-{5 \over 2}), 1), \underline{1}(-3)\Bigr]
+ \Bigl[ \mbox{CPT-conjugate} \Bigr].
\end{eqnarray}

By noting that the helicity of every such state as (3) and (4) is
automatically determined by SO(10) SPA and that
$Q_{1}^{M}$ and  $\overline{Q}_{\dot{1}}^{M}$
satisfy the algebra of the annihilation and the creation operators for the
spin ${1 \over 2}$ particle,
we speculate that these states (3) and (4) are the
reltivistic(gravitational)
massless composite states spanned upon the mathematical(not the physical
vacuum with the lowest energy) Clifford vacuum and are composed of the
fundamental object $Q^{N}$ $\it{superon}$ with spin ${1 \over 2}$.
Therefore we regard (1) as $\it{a}$ ${quintet}$ ${of}$ ${superons}$
and
$\it{a}$ ${quintet}$ ${of}$
${antisuperons}$.
The identification of the generators of SO(10) SPA with the fundamental
objects(particles) is strange so far especially from the viewpoint of the
familiar local gauge field theory.
We will consider these problems later and show (a possibility of) a
fundamental (local) field theory of the superons.

Now we envisage the Planck scale physics as follows:
At(above) the Planck energy scale spacetime and  matter have the
structure described by SO(10) SPA, where the gravity dominates and creates
the superon-quintet and the antisuperon-quintet pair from the vacuum in
such a way as superon-composites massless states span the irreducible
massless
representations of SO(10) SPA.

{}From the viewpoints of the superon hypothesis we can reinvestigate more
concretely the physical meaning of the results obtained in Ref.[3].
For simplicity we use the following notations for superons
$Q^{N}$ ($N=1,2,..10$).

For the superon quintet $\underline 5$:
$\Bigr[ $(\underline 3, \underline 1;-${1 \over 3}$,-${1 \over 3}$,
-${1 \over 3}$)$,
$(\underline 1, \underline 2;1, 0)$ \Bigl], $ we use

\begin{equation}
\Bigr( Q_{a},Q_{b},Q_{c},Q_{m},Q_{n} ;  a,b,c=1,2,3; m,n=4,5\Bigl)
\end{equation}

\noindent and for the antisuperon-quintet $\underline 5^{*}$:
$\Bigr[ (\underline 3^{*}, \underline 1;+{1 \over 3},+{1 \over 3},
+{1 \over 3})$,
$(\underline 1, \underline 2^{*};-1, 0) \Bigl]$, we use

\begin{equation}
\Bigr(Q_{a}^{*},Q_{b}^{*},Q_{c}^{*},Q_{m}^{*},Q_{n}^{*}; a,b,c=1,2,3;
m,n=4,5\Bigl).
\end{equation}

\par

Accordingly  we can specify explicitly all the the states corresponding
to observed quarks, leptons and massless gauge bosons of the
standard model(SM)[4] presented in Ref.[3] as follows.
The multiplicity of the fermionic states identified with respect to
( SU(3), SU(2); electric charges ) is counted in the two-component
Weyl spinor unit. (SO(10) normalization factor is neglected.)
\par

\noindent
\underline{$({\nu}_{e}, e)_{L}$ type leptons}:
four generations from \underline{120}; $(Q_{m}Q_{4}^{*}Q_{5}^{*})$,
$(Q_{a}Q_{a}^{*}Q_{m})$ and conjugate states.

\noindent
four generations from \underline{252};
${\varepsilon}_{abc}Q_{b}Q_{c}
{\varepsilon}_{ade}Q_{d}^{*}Q_{e}^{*}Q_{m}$,$Q_{a}Q_{a}^{*}
{\varepsilon}_{lm}Q_{l}^{*}Q_{m}^{*}Q_{n}$,
and conjugate states.                                              \par

\noindent
\underline{$(e)_{R}$ type leptons}:
two generations from \underline{120};
${\varepsilon}_{abc}Q_{a}Q_{b}Q_{c}$  and  conjugate tstate.

\noindent
two generations from \underline{252};
${\varepsilon}_{abc}Q_{a}Q_{b}Q_{c}Q_{m}Q_{m}^{*}$ and conjugate state.
                                                                    \par
\noindent
\underline{$(u, d)_{L}$ type quarks}:
two generations from \underline{120};
${\varepsilon}_{abc}Q_{b}^{*}Q_{c}^{*}Q_{m}^{*}$ and  conjugate state.

\noindent
four generations from \underline{252};
${\varepsilon}_{abc}Q_{a}^{*}Q_{b}^{*}Q_{c}^{*}Q_{d}Q_{m}^{*}$,
${\varepsilon}_{abc}Q_{b}^{*}Q_{c}^{*}Q_{l}
{\varepsilon}_{mn}Q_{m}^{*}Q_{n}^{*}$ and  conjugate states.
                                                        \par

\noindent
\underline{$(u)_{R}$ type quarks}:
two generations from \underline{120};
$Q_{a}{\varepsilon}_{mn}Q_{m}Q_{n}$  and  conjugate state.

\noindent
two generations from \underline{252};
${\varepsilon}_{abc}Q_{b}Q_{c}Q_{a}^{*}{\varepsilon}_{mn}Q_{m}Q_{n}$
and  conjugate state.                               \par

\noindent
\underline{$(d)_{R}$ type quarks}:
four generations from \underline{120};
$Q_{a}^{*}{\varepsilon}_{abc}Q_{b}Q_{c}$, $Q_{a}Q_{m}Q_{m}^{*}$
and their conjugate states.

\noindent
six generations from \underline{252};
${\varepsilon}_{abc}Q_{a}Q_{b}Q_{c}{\varepsilon}_{def}Q_{e}^{*}Q_{f}^{*}$,${
\varepsilon}_{abc}Q_{b}Q_{c}Q_{a}^{*}Q_{m}Q_{m}^{*}$,
$Q_{a}{\varepsilon}_{kl}Q_{k}Q_{l}{\varepsilon}_{mn}Q_{m}^{*}Q_{n}^{*}$,and
conjugate states.                                             \par

\noindent
\underline{$SU(2) \times U(1)$ gauge bosons}:
one singlet state from \underline{45};
${1 \over \sqrt{2}}( Q_{4}Q_{4}^{*}-Q_{5}Q_{5}^{*})$  and
one triplet states from \underline{45};
\{ $Q_{4}Q_{5}^{*}$, ${1 \over \sqrt{2}}( Q_{4}Q_{4}^{*}+Q_{5}Q_{5}^{*})$,
$Q_{5}Q_{4}^{*} \}$.                                                   \par

\noindent
\underline{$SU(3)$ gluons}:
one octet state from \underline{45};
$\{Q_{1}Q_{3}^{*},Q_{2}Q_{3}^{*},-Q_{1}Q_{2*},{1 \over \sqrt{2}}
(Q_{1}Q_{1}^{*}-Q_{2}Q_{2}^{*})$, $Q_{2}Q_{1}^{*}$,
${1 \over \sqrt{6}}(2Q_{3}Q_{3}^{*}-Q_{2}Q_{2}^{*}-Q_{1}Q_{1}^{*}),
-Q_{3}Q_{2}^{*},Q_{3}Q_{1}^{*}\}$.                                   \par

\noindent
\underline{$SU(2)$ Higgs Boson}:
one doublet state from \underline{210};
${\varepsilon}_{abc}Q_{a}Q_{b}Q_{c}Q_{m}$ and  conjugate state.
                               \par

\noindent
\underline{$( X,Y )$ leptoquark bosons in GUT(SU(5)[5] and SO(10)[6])}:
{}From \underline{45} we obtain:
$Q_{a}^{*}Q_{m}$  and   conjugate state.                            \par

The specifications of SU(5) leptoquark states are interesting
concerning the proton decay, for the symmetry breaking via SU(5) invariance
may be worthwhile to consider.
For the gauge bosons we have considered only the adjoint representation of
SO(10) SPA.

Although the mass generation mechanism, i.e. the mechanism of the symmetry
breaking,

\noindent
[SO(10) SPA(massless)]  \\
$\longrightarrow$ [ ... ] \\
$\longrightarrow$ [ \mbox{Poincar\'e algebra (massless in part)
with} \ $SU(3) \times SU(2) \times U(1)$ ] \\
$\longrightarrow$ [ \mbox{Poincar\'e algebra(massless in part)
with} \ $SU(3) \times U(1)$ ]

\noindent is unknown, we dare to perform $SU(3) \times SU(2) \times U(1)$
invariant recombinations of the helicity states of the massless irreducible
representation of the little algebra of SO(10) SPA in order to see the
possible contents of $SU(3) \times SU(2) \times U(1)$ invariant
massive states of the irreducible representation of the little algebra of
Poincar\'e algebra.                                                 \par
Through the recombination, many of the lower helicity
$( \pm{3 \over 2}, \pm1, \pm{1 \over 2}, 0 )$ states
of SO(10) SPA are converted to the longitudinal components of the higher
spin
massive states of Poinar\'e algebra and others remain massless.

In Ref.[3] we have carried out the recombination among $2 \cdot 2^{10}$
helicity states and
found surprisingly all massless states necessary  and sufficient for
the SM with three generations of quarks and leotons appear in the surviving
massless  states.
A few characteristic(i.e. independent on the intermediate symmetry
breaking pattern) predictions which can be tested by the high energy
particle experiment are presented in Ref.[3].                         \par
Now superons are for quarks, leptons, gauge bosons except the graviton, and
Higgs bosons  what quarks are for baryons and mesons.
Then it is interesting to investigate the
the symmetry breaking (mass pattern) of SO(10) SPA by performing the similar
analysis used in the quark model for the hadron physics[7].
It gives the phenomenological informations to solve the superon dynamics at
the Planck scale.

While, towards the construction of the field theory of
the superon and for surveying the physical(phenomenological) implications of
the superons for the unified gauge models(SM and GUTs) it is very important
to understand all the gauge couplings of the unified gauge models in terms
of
the superon pictures.
For simplicity we neglect the mixing between the states and take by using
the conjugate representations  naively the following left-right symmetric
assignment for quarks and  leptons , i.e.
$( \nu_{l}, \it{l}^{-} )_{R}= (\overline\nu_{l}, \it{l}^{+} )_{L}$, etc.

\noindent
For $({\nu}_{e}, e)$,  $({\nu}_{\mu}, \mu)$,  $({\nu}_{\tau}, \tau)$
we take $(Q_{m}Q_{4}^{*}Q_{5}^{*})$, $(Q_{a}Q_{a}^{*}Q_{m})$,
$Q_{a}Q_{a}^{*}Q_{b}Q_{b}^{*}Q_{m}$
and the conjugate states respectively.
                                            \par
\noindent
For $( u, d )$, $( c, s )$, $( t, b )$
we take  ${\varepsilon}_{abc}Q_{b}^{*}Q_{c}^{*}Q_{m}^{*}$,
${\varepsilon}_{abc}Q_{b}^{*}Q_{c}^{*}Q_{l}
{\varepsilon}_{mn}Q_{m}^{*}Q_{n}^{*}$
${\varepsilon}_{abc}Q_{a}^{*}Q_{b}^{*}Q_{c}^{*}Q_{d}Q_{m}^{*}$,
and the conjugate states respectively.
\par
\noindent
The superon line Feynmann diagrm is obtained by replacing the
single line in the Feynmann diagram of the gauge models by the corresponding
multiple superon lines.
We discuss as a few examples  the following  processes, i.e.\

$(i)$ ${\beta}$ decay:
$n^{0} \longrightarrow p^{+} + e^{-} + \overline{\nu_{e}}$,\quad
$(ii)$ ${\pi}^{0} \longrightarrow 2{\gamma}$,
$(iii)$ the proton decay:
$p^{+} \longrightarrow e^{+} + {\pi}^{0}$,
$(iv)$  a flavor changing neutral current process(FCNC):
$K^{+} \longrightarrow \pi^{+} + \nu_{e} + \overline\nu_{e}$
and
$(v)$ an advocated typical process of the (non-gauged) compositeness:
$\mu \longrightarrow e + \gamma$.
                                                      \par
To translate the vertex of the Feynmann diagram of
the unified gauge model into that of  the superons,
we assume that the superon-antisuperon pair creations and
pair annihilations within a single state for a quark, a lepton and
a (gauge) boson (,i.e. within a single SO(10) SPA state) are forbidden.
This rule seems natural because  every state is the irreducible
representation
of SO(10) SPA and is prohibitted from the decay without any remnants,
i.e.  without the interaction between the superons contained
in the different states.
Now it is straightforward to translate uniquely the Feynmann diagram
of the unified gauge models into that of  the superon model.
                                                       \par
For the processes $(i)$ and $(ii)$ we can draw the corresponding similar
tree-like superon line diagrams easily,
where the triangle-like superon diagram does not appear.
For the process $(iii)$ we consider the Feynmann diagrams for the proton
decay
of GUTs and find that the corresponding superon line diagrams
do not exist due to the mismatch of the superons contained in the
quarks(u and d)  and the gauge bosons(X and Y) at the gauge coupling
vertices.
This means that irrespective of the massses of the gauge bosons
the proton is stable, at least against
$p^{+} \longrightarrow e^{+} + {\pi}^{0}$.
For FCNC process $(iv)$ the penguin-type and the box-type superon line
diagrams are to be studied corresponding to the penguin- and box-Feynmann
diagrams for
$K^{+} \longrightarrow \pi^{+} + \nu_{e} + \overline\nu_{e}$ of GUTs.
Remarkablely the superon line diagrams which have only the up-quark for the
internal quark line exist.
This is the indication of the strong suppression of the FCNC process,
at least for the process
$K^{+} \longrightarrow \pi^{+} + \nu_{e} + \overline\nu_{e}$.
For the process $(v)$ the corresponding tree-like superon line diagram
does not exist, i.e.  $\mu \longrightarrow e + \gamma$ decay mode is
absent in the superon (composite) model.

It should be noticed that among the abovementioned superon line diagrams,
the superon line diagrams corresponding to the penguin Feynmann diagram
have the different topological structures,
i.e. the twist of the superon lines within the propagator occurrs.
Such topological structures remain to be studied in detail.
As an example of the superon line diagram we show in Figure 1
one of the penguin Feynmann diagrams for the process $(iv)$.

Finally we consider the possibility of the fundamental theory which
describes the superon dynamics.
In Ref.[8], we have investigated the nonlinear representation of N=1 SUSY[9]
in two dimensional spacetime.
We have shown that by using Noether procedure we can construct the
supercharge $Q$ explicitly and
carry out the canonical quantization for the fundamental (Goldstone) spinor
field $\psi(x)$ so that the the super-Poincar\'e algebra can be satisfied
at the quantized level.
The spinor supercurrent density giving the supercharge $Q_{\alpha}$ has been
written as follows
\begin{equation}
J^{\mu}(x)={1 \over i}\sigma^{\mu}\psi(x)
-\kappa \{ \mbox{the higher orders of} \ \psi(x) \}, \quad
(\mu=1,2)
\end{equation}
where $\kappa$ is an arbitrary constant with the second power of length and
$\sigma^{\mu}$ are Pauli matrices, $\sigma^{0}=\bf{1}$.
This is a nice indication for our asumption that the generator(supercharge)
$Q^{N}$ of SO(10) SPA represents the fundamental object(particle)
$superon$ $with$ $spin$ ${1 \over 2}$, which obeys the Fermionic quantum
statistics.
(8) is the field-current identiy with spin  ${1 \over 2}$ including,
as shown by the  term proportional to $\kappa$, the composite fields
corresponding to the states of the irreducible representation of SO(10)
SPA generated by the currents.
Therefore we speculate that the fundamental theory of the superon quintet
model of spacetime and matter at(above) the Planck scale is SO(10) nonliner
supersymmetry (NL SUSY) in the curved spacetime,
and all the helicity-states including the observed  quarks,
leptons and gauge bosons except the graviton are the massless relativistic
(gravitational) composite states composed of Goldstone fermions, $superons$.

{}From these considerations
we propose the following Lagrangian as the fudamental theory of
SO(10) superon model of spacetime and matter.

\begin{equation}
L=-{c^{3} \over 16{\pi}G}eR -{c^{3} \over 8{\pi}G}e\Lambda
-{1 \over \kappa}e\vert W \vert,
\end{equation}

\begin{equation}
\vert W \vert=detW_{\mu}^{\nu}=det (\delta_{\mu}^{\nu}+\kappa
T_{\mu}^{\nu}),
\end{equation}
\begin{equation}
T_{\mu}^{\nu}={1 \over
2i}\sum_{i,j=1}^{10}(\overline{s}^{i}O_{ij}\gamma_{\mu}
D^{\nu}{s}^{j}
- D^{\nu}{\overline{s}^{i}}\gamma_{\mu}O_{ij}{s}^{j}),
\end{equation}

where  $\kappa$ is yet an arbitrary constant with the dimension of the
fourth
power of length,  $e=det e^{a}_{\mu}$, $D_{\mu}=\partial_{\mu} +
{1 \over 2}\omega_{\mu}^{ab}\sigma_{ab}$ and $R$ and $\Lambda$ are
the scalar curvature and the yet arbitrary cosmological constant,
respectively.
$O_{ij}$ is a $10 \times 10$ unitary  matrix representing
the quantum mechanical mixing among the superon quintet states,
which may be probable but unpleasant from the elementary nature of the
superon.
The first term of (9) is the gravity action and the third term is
the action for the superon $s(x)^{i},(i=1,2,..,10)$, which is global
SO(10) NL SUSY in the curved spacetime.
The invariance of the total action (9)
under global SO(10) NL SUSY  will be proved explicitly by performing
the 1.5 order formalism adopted for proving the invariance of the
SUGRA action[10].
The states with helicity $\pm3$, $\pm{5 \over 2}$ and
$\pm2$(except the graviton) necessary for completing the irreducible
representation of SO(10) SPA  appear afetr specifying the contorsion in
the spin conection $\omega^{\mu}_{ ab}(e_{a}^{\mu},s^{i})$, for these states
are  10-, 9- and  8-superon states respectively and hidden in (9).
We can anticipate the invariance of (9) under the global SO(10) NL SUSY,
which may be included in the scope of Ref[11].
The structure of the true vacuum of (9) should be studied in detail.

{}From the phenomenological viewpoints, the current algebra
may be useful and practical.
{}From the commutator of the supercharges,

\begin{equation}
\{Q_{\alpha}^{M},\overline{Q}_{\dot{\beta}}^{N}\}
= 2{\delta}^{MN}\sum_{{\mu}
=0}^{3}(\sigma_{\mu})_{{\alpha}{\dot{\beta}}}P^{\mu},
\end{equation}
we can express the Hamiltonian density in the form of the products of the
supercurrents of SO(10) NL SUSY obtained by Noether's procedure as
demonstrated in Ref.[8]. For understanding the superon dynamics
phenomenologically it is important to fit all the decay
data of low lying hadrons in terms of the superon line diagram amplitudes,
which may give interesting structures among the various  amplitudes.

The unified gauge models (SM and GUT) may be for the yet hypothetical
SO(10) superon model what the Landau-Ginzburg theory is for the BCS theory
of
the superconductivity, i.e. they may be the (effective) theories
at the low energy.

Besides those nice aspects of SO(10) superon model mentioned above,
much more open questions are left.

However we speculate that the beautiful complimentality between the gauge
unified models(SM and GUT) and the superon model, i.e. the former is
strengthened or revived by taking account of the topology of the latter
superon diagram, while drawing the superon diagram of the latter is guided
by the Feynmann diagram for the gauge interaction of the former,
may be an evidence of SO(10) SPA structure of spacetime and matter behind
the gauge models, i.e. an evidence of the superon quintet hypothesis .
The experimental search of a predicted new spin ${3 \over 2}$ lepton doublet
$( \nu_{\Gamma}, \Gamma^{-} )$ with the mass of the electroweak scale[3] is
important. The clear signals of the new particles may be similar to
the top-quark pair production event without the jet production.

\vskip 30mm

The author would like to express gratitude to Professor Julius Wess and
Professor Gyo Takeda for their encouragement at the early stage of this
work.
Also he would like to thank Professor Takeshi Shirafuji and Professor
Yoshiaki
Tanii for useful discussions and the hospitality at Physics Department
of Saitama University.
\vskip 10mm

\newpage

%
\newcommand{\NP}[1]{{\it Nucl.\ Phys.\ }{\bf #1}}
\newcommand{\PL}[1]{{\it Phys.\ Lett.\ }{\bf #1}}
\newcommand{\CMP}[1]{{\it Commun.\ Math.\ Phys.\ }{\bf #1}}
\newcommand{\MPL}[1]{{\it Mod.\ Phys.\ Lett.\ }{\bf #1}}
\newcommand{\IJMP}[1]{{\it Int.\ J. Mod.\ Phys.\ }{\bf #1}}
\newcommand{\PR}[1]{{\it Phys.\ Rev.\ }{\bf #1}}
\newcommand{\PRL}[1]{{\it Phys.\ Rev.\ Lett.\ }{\bf #1}}
\newcommand{\PTP}[1]{{\it Prog.\ Theor.\ Phys.\ }{\bf #1}}
\newcommand{\PTPS}[1]{{\it Prog.\ Theor.\ Phys.\ Suppl.\ }{\bf #1}}
\newcommand{\AP}[1]{{\it Ann.\ Phys.\ }{\bf #1}}

\newpage

\bf{Figure captions}: \
\noindent
One of the penguin diagrams of $K^{+}$ decay.
\noindent
Superons are labeled by the indices used in Eqs.(6) and (7).
\noindent
\bf{Figure 1}\
$K^{+} \longrightarrow {\pi}^{+} + {\nu}_{e} + \overline{\nu}_{e}$

\newpage

\bf{Figure 1} \
One of the Penguin diagrams of $K^{+}$ decay

$K^{+} \longrightarrow {\pi}^{+} + {\nu}_{e} + \overline{\nu}_{e}$

\end{document}